\magnification1200


\vskip 2cm
\centerline
{\bf Kac-Moody algebras and the cosmological constant}
\vskip 1cm
\centerline{ Peter West}
\centerline{Department of Mathematics}
\centerline{King's College, London WC2R 2LS, UK}
\vskip 2cm

\vskip1cm
\leftline{\sl Abstract}
We show that the theory of gravity constructed from the non-linear realisation of the semi-direct product of the  Kac-Moody algebra $A_{1}^{+++}$ with its vector representation  does not allow a cosmological constant. The existence of a cosmological constant in this theory is related to the breaking of the gravitational duality symmetry. 
\vskip2cm
\noindent

\vskip .5cm

\vfill
\eject

\medskip
{{\bf 1. Introduction}}
\medskip
Quite some time ago it was conjectured that the underlying  theory of strings and branes had an $E_{11}$ Kac-Moody symmetry [1]. More precisely it was proposed that an effective theory of strings and branes was the non-linear realisation of the semi-direct product of $E_{11}$ with its vector representation, denoted $E_{11} \otimes_s l_1$. This theory has an infinite number of fields which depend on an infinite number of coordinates. However, if one restricts the fields to be those at low levels and takes them to only depend on the usual coordinates of spacetime then the equations of motion are precisely those of the maximal supergravity theories. Strictly speaking this has only been shown in all detail in eleven [2,3], seven [4] and four dimensions [2,3] but it also has to be the case in all dimensions less than eleven. The different theories arise as different decompositions of $E_{11}$. For a review see the references [5] and [6]. 
\par
It was also proposed that the very extended algebra $A_{1}^{+++}$ was associated with gravity [7]. 
Recently the non-linear realisation of the semi-direct product of the very extended Kac-Moody algebra $A_{1}^{+++}$ with its vector representation, denoted $A_{1}^{+++}\otimes_s l_1$ has been constructed [8]. This theory also has an infinite number of fields which depend on an infinite number of coordinates. The equation of motion for the lowest level field when taken to depend on just the lowest level coordinate is just Einstein's theory. The field at the next level is the dual graviton. The Dynkin diagram for $A_{1}^{+++}$ is given by 
$$
\matrix{
\bullet & - & \bullet & - & \bullet & = & \otimes \cr
1 & & 2 & & 3 & & 4 \cr
}
$$

\par
As explained in previous $E_{11}$ papers, and for the case of  $A_{1}^{+++}\otimes_s l_1$,  the field equations are constructed out of the Cartan forms.  Using the vielbein contained in the non-linear realisation we can  converting the form index on the Cartan form   to a tangent index. This object  is invariant under the rigid transformations of the non-linear realisation and just  transforms under the local symmetry. The equations of motion are found by demanding  that they  transform into each other under the symmetries of the non-linear realisation. As the Cartan forms involve a derivative,  one can not find a contribution to the equations of motion that involves no derivatives. This would seem to rule out a cosmological term. 
\par
 While no cosmological term is possible for eleven dimensional supergravity and IIB supergravity. It is possible for IIA supergravity and the maximal supergravities in less than ten dimensions.These theories have become known as gauged supergravities as some of the vector fields also become gauge fields.  In fact   a cosmological term does appear in the non-linear realisation of $E_{11} \otimes_s l_1$ but by a  novel  mechanism which we now explain.  It turns out that the non-linear realisation of $E_{11} \otimes_s l_1$ contains fields that are the so called next to top fields $A_{a_1\ldots a_{D-1}}$, that is, fields in $D$ dimensions which have $D-1$ anti symmetrised indices. The field strength  $F_{a_1\ldots a_{D}}$  for these fields  has  a field  equation that says that this field strength is essentially a constant, that is,  $F_{a_1\ldots a_{D}}= c\epsilon _{a_1\ldots a_{D}}$ where $c$ is a constant. Thus we find what is in effect a cosmological term in the equation of motion.  It would be interesting to examine if this way of introducing a cosmological constant has some phenomological consequences that are different to the more usual mechanism. 
\par
Hence one can find  a cosmological constant in the non-linear realisations mentioned above if there is a next to top form among the fields. We find a theory in  $D$ dimensions in  the $E_{11} \otimes_s l_1$ non-linear realisation by  deleting node $D$ in the $E_{11}$ Dynkin diagram and decomposing the representations into the remaining subalgebra which is $GL(D)\otimes E_{11-D}$. Although there are not any next to top forms in eleven dimensions, or the ten dimensional  theory that contains IIB supergravity, one  does find one  next to top forms in IIA and several next to top forms in all supergravity theories in less than ten dimensions. The next to top forms  in $D$ dimensions belong to representations of $E_{11-D}$ and these are easily found.  The results [9,10] agree with those found previously using the standard supersymmetry of the maximal supergravity theories [11]. Indeed, the representations of $ E_{11-D}$  that the next to top forms belong to agree precisely with the embedding tensor that controls the structure of the gauged supergravities. This mechanism has been demonstrated in detail for the $E_{11} \otimes_s l_1$ non-linear realisation   in ten dimensions [12] where one finds Romans theory [13]. Hence the non-linear realisation of $E_{11} \otimes_s l_1$ contains all maximal supergravity theories and not just the massless supergravity theories. 
\par
To see if we can add a cosmological constant to the theory resulting from the non-linear realisation of $A_{1}^{+++}\otimes_s l_1$ we just have to look for a three form among the fields in the adjoint representation of  $A_{1}^{+++}$ when decomposed to GL(4). Looking at equation (3.2) of reference [8] we find that there is no such field at low levels. In fact all the fields have an even number of indices and so there can be no such field. Hence we conclude that the $A_{1}^{+++}\otimes_s l_1$ symmetry forbids a cosmological constant. 
\par
We can  recover this result from a different view point. It is very well known that if one dimensionally reduces gravity in four  dimensions to three dimensions then one recovers an SL(2,R) symmetry; the Ehlers symmetry [14]. We very briefly recall this much repeated calculation. For example, it  can be carried out using the equations (13.1.1.118- 127) of reference [6].  The vielbein takes the form 
$$
\hat e_{\hat \mu}{}^{\hat a}= \left(\matrix {e^{{\phi\over 2}} e_\mu{}^a & e^{-{\phi\over 2}}A_\mu\cr 
0& e^{-{\phi\over 2}}\cr}\right)
\eqno(1)$$
The field strength of the vector field can be dualised to give a scalar field $\chi$. The results is that the four dimensional Einstein action takes the form 
$$
\int d^4 x (\det \hat e_{\hat \mu}{}^{\hat a}) \hat R= L \int d^3 x   (\det e_\mu{}^a) (R -{1\over 2}{ (\partial_\mu \tau \partial^\mu \tau  )\over (Im \tau )^2})
\eqno(2)$$
where $\int dx^3 =L$, $\tau= \chi+i e^{-\phi}$. As is well known the last term in this action is invariant under the SL(2,R) Ehlers symmetry. 
\par
On the other hand the cosmological term in four  dimensions reduces to three dimensions as 
$$
\int d^4 x  \Lambda (\det \hat e_{\hat \mu}{}^{\hat a}) = L \int d^3 x  \Lambda  e^{\phi} (\det  e_\mu{}^a) 
\eqno(3)$$
which is clearly not SL(2,R) invariant. Hence although when  we find an SL(2,R) Ehlers symmetry when Einstein's  gravity in four dimensions is dimensionally reduced to three dimensions, this  is not the case if we add a cosmological constant in four dimensions. 
\par
The appearance of the Ehlers symmetry can be easily understood from the view point of the non-linear realisation of $A_{1}^{+++}\otimes_s l_1$. Looking at the Dynkin diagram for non-linear realisation of $A_{1}^{+++}$ given above we find the three dimensional theory by deleting node three. The remaining algebra is $GL(3)\otimes SL(2)$. The last factor is the Ehlers symmetry. Hence if it was possible to add a cosmological constant to the 
 non-linear realisation we would find at theory that had the symmetries  of the non-linear realisation and so the SL(2,R) symmetry in three dimensions. This is not the case and so we can not add a cosmological constant. A more straight forward approach is to add the cosmological constant in four dimensions and verify that it is not invariant under the symmetries of the non-linear realisation. 
\par
After Einstein had retracted his introduction of the cosmological constant it was generally believed that the universe had no cosmological constant. However, relatively recently it has become apparent that we now  live in an asymptotically  de sitter space that has a  small cosmological constant. The problem is that this cosmological constant is extremely small compared to any vacuum energy that does arises in quantum field theories such as the standard model. This is one of the most outstanding problems with our current theories. 
\par
One intriguing fact which is at first sight relevant to this problem concerns supersymmetric theories. It was realised early on in the development of supersymmetry that if supersymmetry was preserved then the effective potential had no quantum corrections [14]. This meant that in a rigid theory of supersymmetry if there is no vacuum energy at the classical level then there will be none in the quantum theory. The problem with approach  is that in a realistic theory of nature   supersymmetry has to be  broken at a relatively high scale and so this protection disappears and one generically  finds a very large cosmological constant. 
\par
The non-linear realisation of $A_{1}^{+++}\otimes_s l_1$ contains an infinite number of fields. Many of these are dual descriptions of gravity, but there are also  fields whose physical meaning is as yet  unknown. However, it is suspected that it does not contain fields that lead to propagating degrees of freedom in addition to the field of gravity. As such it is a theory of gravity which is very closely related to what we observe. To have a cosmological constant the $A_{1}^{+++}$ would have to be broken, but unlike for supersymmetry one might manage to do this at a very low scale. The lowest level transformation in $A_{1}^{+++}$, apart from the GL(4) symmetry at level zero,  has the parameter $\Lambda ^{ab}= \Lambda ^{(ab)}$ and it transforms the graviton into the dual graviton and visa-versa. Indeed these two transformations generate the entire $A_{1}^{+++}$ algebra. Hence, from this viewpoint,  we find that the breaking of the gravity-dual gravity symmetry is  related to the appearance of a cosmological constant. While this suggestion is far from a solution of the cosmological constant puzzle it does introduce some connections between preciously unrelated subjects. 
It would be interesting to investigate this possibility in more detail. The original use of non-linear realisations was to account for pion dynamics. In the non-linear realisation the pions were massless but they were given a mass by hand. Adding the cosmological constant would be the analogue of this mass. 
\par
The predictions of $E_{11}$ are sometimes similar to those that result from the supersymmetry algebra. One of these facts  is that one can not add a cosmological constant to the eleven dimensional supergravity theory. One can construct other theories by considering the non-linear realisation of the very extended algebras. It would be interesting to examine which of these theories have a cosmological constant. The general pattern may be that if one takes the theory in its maximum dimension as regards the usual coordinates of spacetime then one finds no cosmological constant is possible. 
However, by deleting a different node in the Dynkin diagram to find a theory in a smaller dimension one can find a cosmological constant.  This is the pattern with $E_{11}$. The advantage with the non-linear realisation considered in this paper is that it involves just gravity and so it is closely  related to what we see.


\medskip
{\bf {Acknowledgments}}
\medskip
I wish to thank Keith Glennon  and Dionysios Anninos  for discussions and the SFTC for support from Consolidated grants number ST/J002798/1 and ST/P000258/1.

\medskip
{\bf {References }}
\medskip
\item{[1]} P. West, {\it $E_{11}$ and M Theory}, Class. Quant. Grav.  {\bf 18}, (2001) 4443, hep-th/ 0104081.
\item{[2]} A. Tumanov and P. West, {\it E11 must be a symmetry of strings and branes},  arXiv:1512.01644.
\item{[3]} A. Tumanov and P. West, {\it E11 in 11D}, Phys.Lett. B758 (2016) 278, arXiv:1601.03974.
\item{[4]} M. Pettit and P. West,  {\it E theory in seven dimensions}, Int.J.Mod.Phys. A34 (2019) no.25, 1950135,  arXiv:1905.07330. 
\item{[5]} P. West,{\it  A brief review of E theory}, Proceedings of Abdus Salam's 90th  Birthday meeting, 25-28 January 2016, NTU, Singapore, Editors L. Brink, M. Duff and K. Phua, World Scientific Publishing and IJMPA, {\bf Vol 31}, No 26 (2016) 1630043, arXiv:1609.06863. 
\item{[6]} P. West, {\it Introduction to Strings and Branes}, Cambridge University Press, 2012.
\item{[7]}  N. Lambert and P. West, {\it Coset Symmetries in Dimensionally Reduced Bosonic String Theory}, Nucl.Phys. B615 (2001) 117-132, hep-th/0107209. 
\item{[8]} K. Glennon and P. West, {\it Gravity, Dual Gravity and A1+++ }, arXiv:2004.03363. 
\item{[9]} F. ÊRiccioni and P. West, {\it The $E_{11}$ origin of all maximal supergravities}, ÊJHEP {\bf 0707} (2007) 063; ÊarXiv:0705.0752.
\item{[10]} E. Bergshoeff, I. De Baetselier and  T. Nutma, {\it  E(11) and the Embedding Tensor},  JHEP 0709 (2007) 047, arXiv:0705.1304. 
\item{[11]} See for example, B. de Wit, H. Nicolai and H.Samtleben, {\it Gauged Supergravities, Tensor Hierarchies, and M-Theory,} JHEP {\bf 0802} (2008) 044, arXiv:0801.1294 [hep-th]; B. de Wit and H. Samtleben, {\it Gauged maximal supergravities and hierarchies of nonabelian vector-tensor
  systems,} Fortsch.\ Phys.\  {\bf 53} (2005) 442, arXiv:hep-th/0501243.
 \item{[12]} A. Tumanov and and P. West, {\it $E_{11}$,  Romans theory and higher level duality relations}, IJMPA, {\bf Vol 32}, No 26 (2017) 1750023,  arXiv:1611.03369
 \item{[13]} L. Romans, {\it Massive $N = 2A$ supergravity in ten dimensions}, Phys. Lett. B169 (1986) 374.
 \item{[14]} J. Ehlers, {\it Transformations of static exterior solutions of EinsteinÕs gravitational field equations into different solutions by means of conformal mapping},  Colloq. Int. CNRS 91, 275 (1962).

\end